\documentclass[aps,prl,reprint,showpacs,preprintnumbers,superscriptaddress]{revtex4-1}
\pdfoutput=1

\makeatletter
\DeclareRobustCommand*\textsubscript[1]{%
  \@textsubscript{\selectfont#1}}
\def\@textsubscript#1{%
  {\m@th\ensuremath{_{\mbox{\fontsize\sf@size\z@#1}}}}}
\makeatother

\usepackage[utf8]{inputenc}
\usepackage{xspace,miller}
\usepackage[separate-uncertainty = true, range-phrase = --, range-units = single]{siunitx}
\usepackage[version=3]{mhchem}

\newcommand{\eg}{e\textsubscript{g}\xspace}
\newcommand{\ttg}{t\textsubscript{2g}\xspace}
\newcommand{\LL}[1]{L\textsubscript{#1}\xspace}

\begin{document}

\title{Real-space mapping of electronic orbitals}

\author{Stefan Löffler}
\email{stefan.loeffler@tuwien.ac.at}
\affiliation{Department for Materials Science and Engineering, McMaster University, 1280 Main Street West, L8S 4M1 Hamilton, Ontario, Canada.}
\affiliation{University Service Centre for Transmission Electron Microscopy, TU Vienna, Wiedner Hauptstraße 8–10/E057B, 1040 Wien, Austria.}
\affiliation{Institute for Solid State Physics, TU Vienna, Wiedner Hauptstraße 8–10/E138, 1040 Wien, Austria.}

\author{Matthieu Bugnet}
\affiliation{Department for Materials Science and Engineering, McMaster University, 1280 Main Street West, L8S 4M1 Hamilton, Ontario, Canada.}

\author{Nicolas Gauquelin}
\affiliation{Department for Materials Science and Engineering, McMaster University, 1280 Main Street West, L8S 4M1 Hamilton, Ontario, Canada.}

\author{Sorin Lazar}
\affiliation{FEI Electron Optics, Achtseweg Noord 5, 5651 GG Eindhoven, The Netherlands.}

\author{Elias Assmann}
\affiliation{Institute for Solid State Physics, TU Vienna, Wiedner Hauptstraße 8–10/E138, 1040 Wien, Austria.}

\author{Karsten Held}
\affiliation{Institute for Solid State Physics, TU Vienna, Wiedner Hauptstraße 8–10/E138, 1040 Wien, Austria.}

\author{Gianluigi A. Botton}
\affiliation{Department for Materials Science and Engineering, McMaster University, 1280 Main Street West, L8S 4M1 Hamilton, Ontario, Canada.}

\author{Peter Schattschneider}
\affiliation{University Service Centre for Transmission Electron Microscopy, TU Vienna, Wiedner Hauptstraße 8–10/E057B, 1040 Wien, Austria.}
\affiliation{Institute for Solid State Physics, TU Vienna, Wiedner Hauptstraße 8–10/E138, 1040 Wien, Austria.}

\begin{abstract}
Electronic states are responsible for most material properties, including chemical bonds, electrical and thermal conductivity, as well as optical and magnetic properties. Experimentally, however, they remain mostly elusive. Here, we report the real-space mapping of selected transitions between p and d states on the Ångström scale in bulk rutile (\ce{TiO2}) using electron energy-loss spectrometry (EELS), revealing information on individual bonds between atoms. On the one hand, this enables the experimental verification of theoretical predictions about electronic states. On the other hand, it paves the way for directly investigating electronic states under conditions that are at the limit of the current capabilities of numerical simulations such as, e.g., the electronic states at defects, interfaces, and quantum dots.
\end{abstract}

\maketitle

Electronic states shape the world around us as their characteristics give rise to nearly all macroscopical properties of materials. Be it optical properties such as colour and refractive index, chemical properties such as bonding and valency, mechanical properties such as adhesion, strength and ductility, electromagnetic properties such as conductance and magnetisation, or the properties of trap states: ultimately, all these properties can be traced back to the electronic states in the material under investigation. Therefore, it is not surprising that electronic states are of paramount importance across many fields, including physics, materials science, chemistry and the life sciences. It does come as a surprise, however, that while some of their aspects can be inferred indirectly from macroscopical material properties or some diffraction techniques, the direct observation of individual electronic states in real space so far has succeeded only under very special circumstances (e.g. on an insulating surface using a scanning tunnelling microscope (STM) with a specially functionalised tip \cite{S_v312_i5777_p1196}) due to both experimental and theoretical challenges. In this work, we endeavour to remedy this situation by using a combination of transmission electron microscopy (TEM), electron energy-loss spectrometry (EELS), and state-of-the-art simulations.

TEM is a well-known technique for studying materials on the nanoscale while EELS adds element-specific information. Both are widely-used on a regular basis in many fields and are readily commercially available. Owing to these two techniques, tremendous progress has been made over the last decade in mapping atom positions with $\approx\SI{10}{\pico\meter}$ accuracy \cite{Urban2008506,N_v423_i6937_p270,NM_v8_i4_p260}, determining charge densities \cite{N_v401_i6748_p49,NM_v10_i3_p209,nakashima2011bonding}, and performing atom-by-atom chemical mapping \cite{N_v450_i7170_p702,N_v468_i7327_p10881090,PRL_v99_i8_p86102,S_v290_i5500_p2280,S_v319_i5866_p1073}. Furthermore, the fine-structures of the spectra allow the determination of the local chemical and structural environment as well as the hybridisation state of the scattering atoms \cite{S_v290_i5500_p2280,N_v490_i7420_p384,S_v319_i5866_p1073,PRL_v107_i_p107602,PRB_v88_i_p115120,ZhouPRL109,PRL_v85_i_p1847,PRB_LaNiO3LaAlO3} in the bulk, which can be substantially different from the surface states probed by STM. This suggests to use the EELS signal to probe the local environment in real-space and map, e.g., crystal fields, conduction states, bonds, and orbitals. Recently, it has been shown on theoretical grounds \cite{U_v131_i0_p39} that such real-space mapping of transitions between orbitals on the Ångström scale should indeed be possible, even though the experimental realisation was expected to be extremely challenging.

\begin{figure*}
	\includegraphics{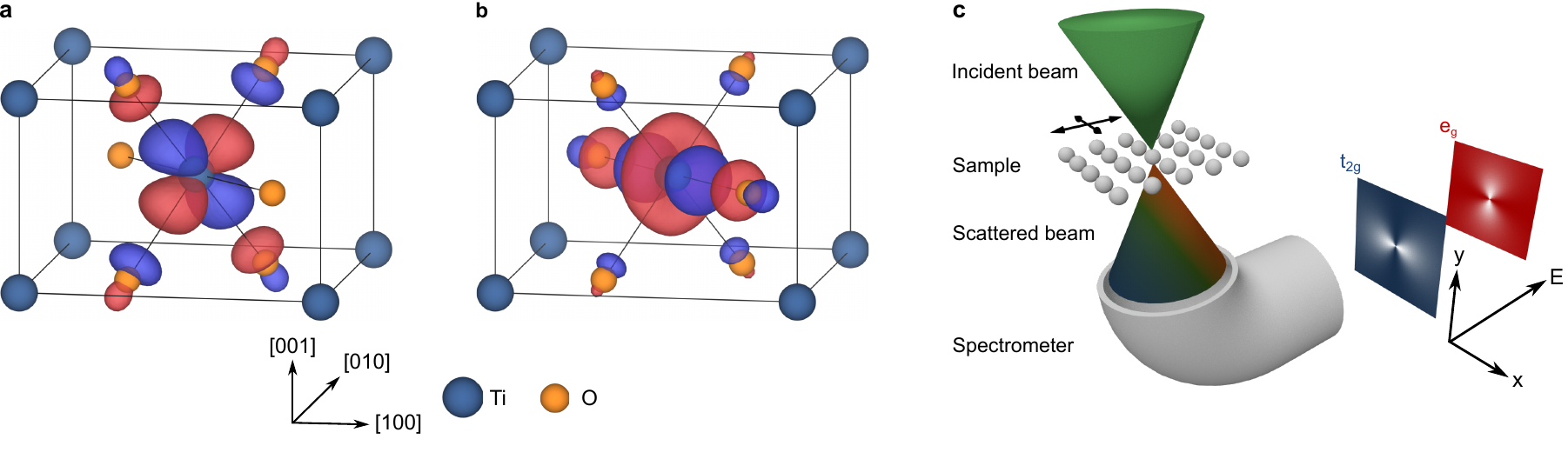}
	\caption{Maximally-localized Wannier functions in rutile corresponding to unoccupied orbitals of \ce{Ti}-\eg character ($x^2-y^2$-like Wannier function (a), $z^2$-like Wannier function (b)). (c) Sketch of the measurement setup. The incident beam is focused onto and scanned over the sample. It can exchange energy and momentum with the specimen, leading to a mixture of states in the scattered beam. Using a spectrometer comprised of a sector magnet and a subsequent imaging system, maps can be formed of all electrons that have transferred a certain amount of energy $E$ corresponding to transitions to different unoccupied orbitals inside the sample. The blue and red planes symbolize the real-space distribution of the transition probabilities to different final states.}
	\label{fig:setup}
\end{figure*}

The method of choice to demonstrate the possibility of this real-space mapping used throughout this work is high-resolution scanning TEM (STEM) together with EELS. In STEM, an electron beam is typically produced by a high-brightness field-emission gun, accelerated to a kinetic energy of the order of \SI{100}{\kilo\electronvolt}, and subsequently focused to an Ångström-sized spot on the sample (see Fig. \ref{fig:setup}c and \cite{WilliamsCarter1996}). Inside the specimen, the probe electrons scatter off the nuclei and sample electrons via the Coulomb interaction.

Scattering by the nuclei is predominantly elastic, i.e., only momentum but no energy is transferred from the lattice (which is assumed to be infinitely heavy) to the probe electron. This gives rise to atomic column contrast in high-resolution TEM, as well as to channeling and dechanneling effects in samples that are thicker than a few tens of nanometres \cite{U_v96_i3--4_p251,U_v96_i3-4_p343}. Dechanneling, which can be visualised as a “hopping” of the electron beam between adjacent columns, destroys the direct spatial correlation between the measured scattering intensity and its point of origin. Consequently, very thin specimens, as well as simulations taking elastic scattering into account, are needed to reduce artefacts and arrive at a reliable interpretation of the data.

Here, the interaction of primary interest is the inelastic scattering of the probe electrons on the sample electrons. Both energy and momentum can be transferred between the beam and the sample. Of particular importance for the real-space mapping of electronic transitions is the so-called core-loss regime of energy transfers of $\gtrsim\SI{100}{\electronvolt}$. They trigger an excitation of a sample electron from an initial, occupied core state to a final, unoccupied conduction-band state. The initial states are typically localised in close proximity to the nucleus and are characterised by a large binding energy. Therefore, crystal-field effects are mostly negligible for core states, which typically exhibit atomic character. The final states, on the other hand, lie close to the Fermi energy, and are strongly influenced by the local environment (see Fig. \ref{fig:setup}a, b).

Due to the strong localisation of the probe beam, it is possible to map the position and energy-dependent transition matrix elements between the initial and the final states using STEM-EELS (see Fig. \ref{fig:setup}c). Given the initial state, it is furthermore possible to obtain both the angular and the radial dependence of the final states \cite{U_v131_i0_p39,U_v111_i_p1163,Loeffler2013} and, thus, bonding information on individual atomic columns \cite{PRB_v88_i_p115120,M_v63_i_p15}. To that end, specific transitions can be selected by using a sufficiently narrow energy range.


\begin{figure}
	\centering\includegraphics{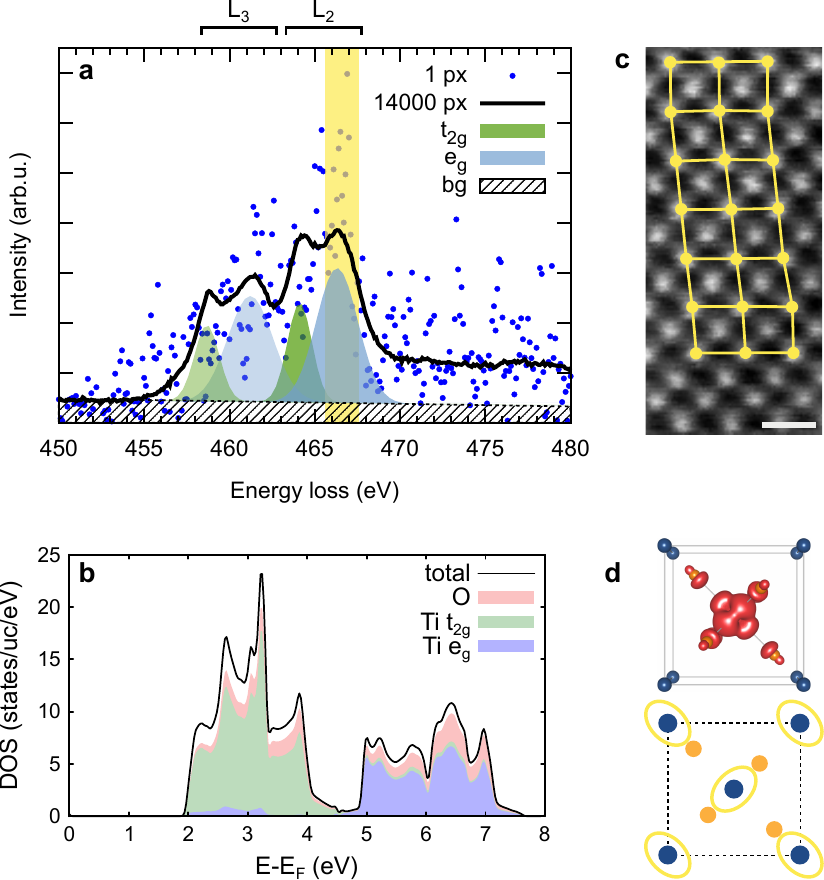}
	\caption{(a) \ce{Ti} \LL{2,3} edge extracted from a single pixel (dots), and averaged over all 14000 pixels (line) of the data set. The energy window used for the energy-filtered \eg maps is highlighted in yellow. Gaussian least squares fits representing the individual shapes of the \eg and \ttg contributions are depicted in blue and green. (b) Projected DOS above the Fermi energy $E_F$ at the position of the \ce{Ti} atoms as calculated by WIEN2k. (c) Dark field image acquired simultaneously with the spectrum image dataset. The spatial distortion, highlighted with yellow lines, is corrected and the unit cells are subsequently averaged. The scale bar indicates \SI{5}{\angstrom}. (d) Unit cell along the \hkl[0 0 1] direction used in the experiment with the summed three-dimensional charge density of the \eg Wannier functions in Fig. \ref{fig:setup}a, b. Also shown are the projected positions of \ce{Ti} (blue) and \ce{O} (orange) atoms as well as yellow ellipses indicating the nearest \ce{O} neighbours of each \ce{Ti} (due to the projection, only two of the four nearest neighbours are visible).}
	\label{fig:SI}
\end{figure}

As a model system, we have chosen rutile (\ce{TiO2}). It has a relatively simple, tetragonal unit cell and, together with the other titanium oxides, has great practical importance, e.g., in renewable energy and energy storage applications, photocatalysis, or as coating material (for a review, see \cite{CR_v114_i19_p9487} and other articles published in the same issue). Its tetragonal structure leads to a strong crystal-field splitting. In particular, the different \ce{Ti-O} bond lengths give rise to a strong asymmetry and splitting \cite{ACSA_v59_i4_p341} of the \eg and \ttg states. Most noticeably, the asymmetric shape of the orbitals is rotated by \SI{90}{\degree} for adjacent \ce{Ti} atoms due to the crystal symmetry (see Fig. \ref{fig:SI}). We concentrate here solely on mapping the \eg states since the \ttg peak has a much lower intensity. Throughout its narrow energy range \cite{NP_v6_i1_p25}, there is always a sizeable \eg contribution (see Fig. \ref{fig:SI}a), making it impossible to identify an unequivocal \ttg signal with today’s instruments due to signal-to-noise ratio (SNR) limitations. Also note that in our simulations, the \ttg Wannier states are much more localized around the nucleus, which strongly reduces the asymmetry caused by crystal-field effects for \ttg states.

\begin{figure}
	\centering\includegraphics{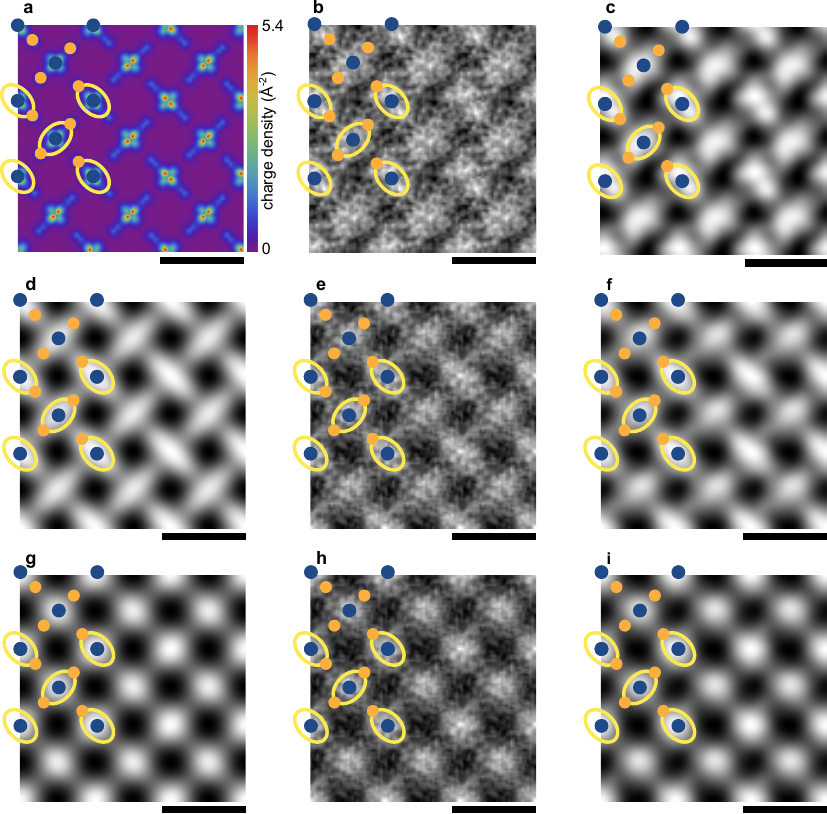}
	\caption{(a) Charge density for the unoccupied \eg orbitals projected along the \hkl[0 0 1] crystallographic axis as calculated by WIEN2k. (b) Experimental energy-filtered map for the \ce{Ti} L ionization edge for final states with \eg character after unit-cell averaging. (c) Same as (b), but after Gaussian smoothing. (d) Simulated energy-filtered map using the multislice algorithm and the MDFF approach after Gaussian blurring. (e) Same as (d) with added noise to better mimic the experimental conditions. (f) Same as (e) after Gaussian smoothing. (g) Simulated energy-filtered map assuming independent atoms without bonding information after Gaussian blurring. (h) Same as (g) with added noise to more closely resemble the experimental conditions. (i) Same as (h) after Gaussian smoothing. All maps are replicated in a $3 \times 3$ raster for better visibility. Overlays show the projected positions of \ce{Ti} (blue) and \ce{O} (orange) atoms as well as yellow ellipses indicating the nearest \ce{O} neighbours of each \ce{Ti} (due to the projection, only two of the four nearest neighbours are visible). All scale bars indicate \SI{5}{\angstrom}.}
	\label{fig:exp}
\end{figure}

A rutile single crystalline sample (MTI corporation) was mechanically thinned down to electron transparency by using the wedge polishing technique with a Multiprep polishing apparatus (Allied High Tech Products Inc.). Further ion milling with a Gentle Mill (Technoorg Linda Ltd.) was performed for ion beam energies in the range of \SIrange{500}{900}{\electronvolt} to remove the damaged regions from the mechanical polishing and provide large, thin, and clean surfaces.

The experiments were performed at \SI{80}{\kilo\volt} acceleration voltage on a FEI Titan 80--300 TEM equipped with spherical aberration correctors and a Gatan GIF Quantum Energy Filter. During the experiment, the single-crystalline sample was oriented in \hkl[0 0 1] direction and the thickness was determined to be \SI{20}{\nano\meter} using EELS \cite{JoEMT_v8_i2_p193,Egerton1996}. A spectrum image (SI) data cube (see Fig. \ref{fig:SI}a) was recorded over several unit cells, together with the elastic dark-field (DF) signal (see Fig. \ref{fig:SI}c). The SI data and the DF signal were acquired simultaneously with a convergence semi-angle of \SI{19}{\milli\radian}, a GIF collection semi-angle of \SI{20.7}{\milli\radian}, and a pixel dwell time of \SI{5}{\milli\second} to maximize the signal while minimizing drift and beam damage artefacts. Optimising the acquisition conditions is essential for acquiring data with sufficient spatial and energy resolution, as well as sufficient SNR for the subsequent data analysis.

The residual lateral drift \cite{U_v106_i11-12_p1129} was corrected using the DF data and the resulting data cube was averaged over 12 unit cells to improve the SNR. Finally, the map corresponding to \LL2 transitions with energy transfers in the range \SI{465 +- 1}{\electronvolt} was extracted. This corresponds to transitions from initial states with $2\text{p}_{1/2}$ character to final states with an energy in the range of \SI{6 +- 1}{\electronvolt} above the Fermi energy, which have mainly \eg character (see Fig. \ref{fig:SI}b and \cite{NP_v6_i1_p25}). Fig. \ref{fig:exp}b shows the resulting energy-filtered map, while Fig. \ref{fig:exp}c shows the same map after Gaussian smoothing. The asymmetry around each \ce{Ti} column is clearly visible, as is the expected \SI{90}{\degree} rotation between nearest neighbours, owing to the different electronic environment caused by the \ce{Ti-O} bonds.

As stated above, comparison to theory is indispensable for a reliable interpretation. To check the results, we simulated the energy-filtered image for the selected energy range and the experimental parameters. To that end, we used the mixed dynamic form factor (MDFF) approach \cite{Schattschneider1986,U_v131_i0_p39,Loeffler2013} based on density functional theory data obtained from WIEN2k\footnote{The simulations were done based on WIEN2k \cite{Wien2k} calculations with the PBE-GGA \cite{PRL_v77_i18_p3865} exchange-correlation potential. Maximally-localised Wannier functions were computed from the \eg bands using the wien2wannier \cite{CPC_v181_i_p1888} and Wannier90 \cite{CPC_v178_i_p685} packages and the disentanglement procedure to separate the target bands from the \ttg band which crosses them near $\Gamma$.} \cite{Wien2k} to model the inelastic interaction between the probe beam and the sample electrons, while the elastic scattering before and after the inelastic scattering event was taken care of using the multislice algorithm \cite{AC_v10_i10_p609,Kirkland1998}. The resulting maps were blurred using a Gaussian filter to account for the finite source size in the experiment. Moreover, noise equivalent to the experimental condition was added to facilitate a visual comparison with the measured data. The resulting simulated map is shown in Fig. \ref{fig:exp}e. The map after Gaussian smoothing, with the same parameters as in Fig. \ref{fig:exp}c, is shown in Fig. \ref{fig:exp}f, where the intensity variations between nearest neighbours are a consequence of the added noise.



The resulting simulated images (see Figs. \ref{fig:exp}e and \ref{fig:exp}f) are found to be in very good agreement with the experimental data (see Figs. \ref{fig:exp}b and \ref{fig:exp}c). In particular, the intensity distribution around the \ce{Ti} atoms is not circular but has a distinct asymmetry towards the nearest \ce{O} atoms (those which lie in the plane of the $x^2-y^2$ Wannier orbital in Fig. \ref{fig:setup}a). In a cubic crystal, there would be no such asymmetry between the \hkl[1 1 0] and \hkl[1 -1 0] directions. Therefore, the experimental data shows the preferential bond direction towards the nearest \ce{O} atoms. This is confirmed by the calculated charge density of the \eg conduction states as depicted in Fig. \ref{fig:exp}a which exhibits the same asymmetry as the experimental map.

To assess the influence of elastic scattering, we also performed calculations with the same elastic scattering potential, but no bonding information. For this purpose, we assumed independent, spherically symmetric atoms without crystal-field effects for the calculation of the inelastic scattering. The resulting data was processed as before to ensure comparability. The resulting maps are shown in Figs. \ref{fig:exp}g--i. In contrast to the \eg maps in Figs. \ref{fig:exp}d--f, the independent-atoms maps in Figs. \ref{fig:exp}g--i cannot reproduce the distinct asymmetry found in the measured data. Thus, it can be concluded that while elastic scattering does affect the signal in principle, under the conditions used in this work elastic scattering alone is insufficient to describe the experimental \eg maps (Figs. \ref{fig:exp}b--c). As they can only be reproduced when taking into account inelastic scattering on realistic electronic states, the peculiar asymmetric shapes can clearly be attributed to the orbital shapes.

In this work, we have demonstrated that the real-space mapping of electronic transitions to specific orbitals is possible in a high-end TEM using EELS, thereby revealing information about both the electronic states themselves and the bonds between atoms. This method --- together with soon realisable improvements in the SNR and accompanied by simulations --- opens the road to studying electronic states in real-space, such as defect states at bulk grain boundaries, bonds at interfaces, or confined electron waves in quantum dots.




St.L. thanks Walid Hetaba for discussions about WIEN2k. St.L. and P.S. thank Ralf Hambach and Ute Kaiser for many valuable discussions. M.B. thanks Vienna University of Technology for travel support. St.L. and P.S. acknowledge financial support by the Austrian Science Fund (FWF) under grant number I543-N20, SFB F45 FOXSI; St.L. also acknowledges financial support by the Austrian Science Fund (FWF) under grant number J3732-N27; E.A. and K.H. by the European Research Council under the European Union’s seventh framework programme (FF 2007/2013)/ERC through grant agreement no. 306447. M.B., N.G., S.L. and G.A.B. performed the experimental work at the Canadian Center for Electron Microscopy, a national facility supported by McMaster University and the Natural Sciences and Engineering Research Council of Canada (NSERC). G.A.B. is grateful to NSERC for supporting this work.


\bibliography{mypubs,papers}

\end{document}